# International Journal of Aerospace Engineering

# A Low-complexity Noncoherent Maximum Likelihood Sequence Detection Scheme for CPM in Aeronautical Telemetry

You Zhou,[1] Rongke Liu,[1] Ruifeng Duan,[2] and Bofeng Jiang[3]

[1] School of Electronic and Information Engineering, Beihang University, Beijing 100191, China

[2] School of Information Science and Technology, Beijing Forestry University, Beijing 100083, China

[3] Beijing Xinwei Telecom Technology CO, LTD, Beijing 100193, China

Correspondence should be addressed to You Zhou; youzhou2015@163.com

## Abstract

Due to high spectral efficiency and power efficiency, the continuous phase modulation (CPM) technique with constant envelope is widely used in aeronautical telemetry in strategic weapons and rockets, which are essential for national defence and aeronautic application. How to improve the bit error rate (BER) performance of CPM and keep a reasonable complexity is key for the entire telemetry system and has been the focus of research and engineering design. In this paper, a low-complexity noncoherent maximum likelihood sequence detection (MLSD) scheme for CPM is proposed. In the proposed method, the criterion of noncoherent MLSD for CPM is derived when the carrier phase is unknown, and then a novel Viterbi algorithm (VA) with modified state vector is designed to simplify the implementation of noncoherent MLSD. Both analysis and experimental results show that the proposed approach has lower computational complexity and does not need accurate carrier phase recovery, which overcomes the shortage of traditional MLSD method. What's more, compared to the traditional MLSD method, the proposed method also achieves almost the same detection performance.

## Introduction

The continuous phase modulation (CPM) signal has constant envelop which makes it very attractive. It can be amplified by nonlinear power amplifiers therefore it has very high power efficiency. Besides, the CPM signal also has high spectral efficiency. What's more, when noncoherent demodulation method is used, the CPM signal is resistant to flame interference and phase interference [1]. These characteristics make it suitable for power limited and high dynamic applications, such as aeronautical telemetry. In current telemetry systems, the pulse code modulation/frequency modulation (PCM/FM), one kind of CPM, has been widely adopted as a main technique ever since 1970s [2]. However, with the increase of the transmission distance and the growing of telemetry data, the limitation of traditional detection method is becoming increasingly obvious especially for rare spectrum resources. The advanced ranging telemetry (ARTM) is working to solve this problem by developing advanced modulation schemes [3], such as FQPSK [4] and ARTM CPM [5]. These schemes achieve higher spectral efficiency than PCM/FM while maintaining the same detection efficiency of PCM/FM at the same time. In this paper, considering the difference between CPM and FQPSK and the application of aeronautical telemetry, we focus on PCM/FM and ARTM CPM which have unified signal representation.

How to improve the detection performance of the aeronautical telemetry system has always been a key focus of research [6-10]. The multi-symbol detection (MSD) is proved to be an efficient way to improve the detection performance of PCM/FM at present. It was first proposed by Osborn and Luntz in 1974 to detect CPFSK [11]. In fact, the PCM/FM signal can be seen as a special case of CPFSK signal, and MSD is extended to the detection of PCM/FM by Geoghegan in 2000 [12]. In 2009 a BQCR-MSD method was proposed [13], and to some extent it reduced the computational complexity of MSD. And then an optimal design of MSD for FM demodulator based on FPGA was explored in [14]. Moreover, the design and evaluation for the MSD algorithm based on GPU is addressed in [15]. The MSD observes several symbol intervals and compares all possible waveforms to make decision for the middle bit. And the detection performance will be better when observing more symbol intervals. However, the computational complexity will also increase exponentially in terms of the number of symbol intervals. Subject to the detection complexity, the MSD method in practical application usually observes no more than 5 symbols, which makes the optimal performance cannot be achieved.

For ARTM CPM, maximum likelihood sequence detection (MLSD) is used to achieve better performance. The MSD method is under the criterion that minimizes the error probability of the decision bits. By contrast, through treating the whole sequence as entirety, the MLSD method minimizes the error probability of the whole sequence instead of the decision bits. The MLSD method can also be applied to PCM/FM to improve its detection performance. But it is too complicate to be realized easily for practical implementation. For example, in PCM/FM when the VA is used to implement MLSD, there are 80 states in total, and 160 branch metrics need to be updated when receiving one new symbol. Similarly, in ARTM CPM there are 512 states in total, and it needs to update 2048 branch metrics.

On the other hand, some methods have also been developed to reduce the complexity of MLSD. In [6], Aulin, Sundberg and Svensson proposed the frequency pulse truncation method to reduce the detection complexity. A state space partitioning method was proposed by Larsson to reduce the complexity of detecting the CPM signal in [16]. In [17-18], it is shown that the matched filters can be reduced by decomposing ARTM CPM into a series of pulse amplitude modulation (PAM) waveforms or a set of orthogonal basis functions. All these works were well summarized by Perrins and Rice in [19]. They made a comparison for several methods in terms of performance and complexity, and pointed out that proper combination of two or more complexity-reducing methods outperforms a single method in terms of detection performance. However, the combination of methods cannot be realized easily in practical system because of the increase of complexity.

What's more, when the MLSD method is used, it is under the hypothesis that the timing synchronization and frequency synchronization have been finished and the phase of carrier must be known accurately, which is always difficult in practical applications [20-21]. Consequently, the noncoherent MLSD method is more valuable for practical application. In 2002, the noncoherent MLSD for single-h CPM was proposed by Colavolpe and Raheli [7]. When it is directly applied to ARTM CPM, considerable principal pulses are needed to maintain the reasonable performance, which leads to high complexity and makes the implementation difficult. In addition, PSP method is proposed to solve detection problem when the phase of the received signal is unknown [22]. But the PSP introduces extra complexity to estimate uncertain phase based on original trellis and the virtual modulation index cannot be obtained easily for ARTM CPM.

In this paper, we explore a low-complexity noncoherent MLSD method for PCM/FM and ARTM PCM, and design a simplified implementation structure based on the analysis for the state transition. In the proposed scheme a new state vector for VA is defined so that the number of states decreases significantly, leading to low complexity and keeping almost the same detection performance. This result is also validated by several simulations, showing the superiority of the proposed scheme.

The reminder of this paper is organized as follows. First we introduce the system model and the MLSD method, and then the low-complexity noncoherent MLSD method is presented. Next, the analysis of complexity and the experiment results are provided, followed by some concluding remarks.

## System model and maximum likelihood sequence detection

### System model

The system block diagram is shown in Figure 1. At the transmitter, the CPM signal is generated as follows. First, the non-return-zero (NRZ) source data is smoothed through a low-pass filter (Pre-filter in Figure 1), a six-order Bessel (Figure 2) for PCM/FM or raised cosine filter (Figure 3) for ARTM CPM, then the signal after filtering is delivered to the FM modulator.

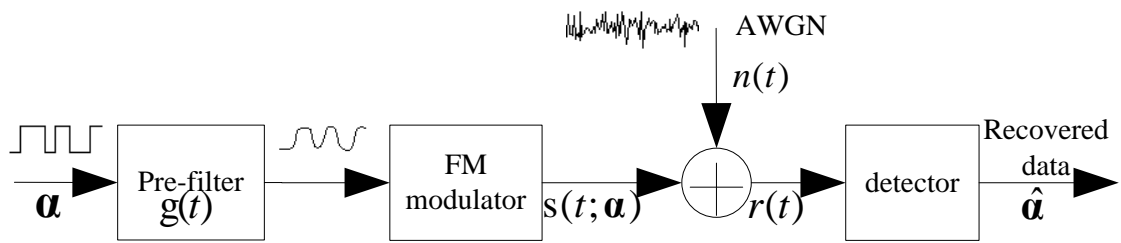

Figure 1: System model of CPM.

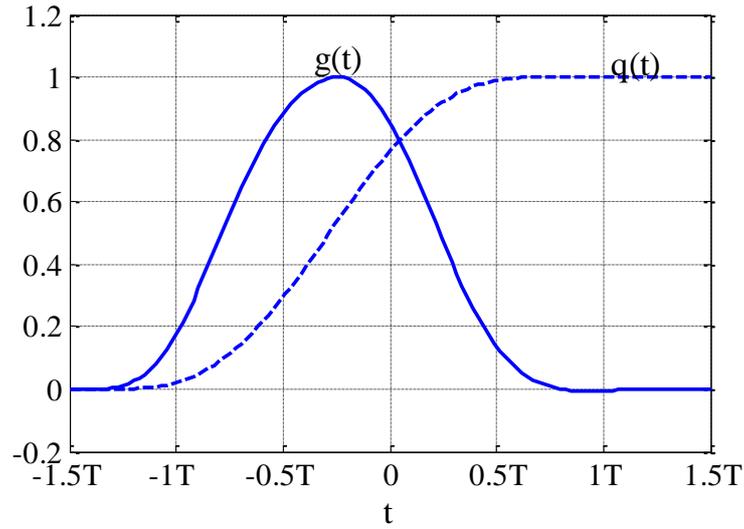

Figure 2: frequency pulse and corresponding phase pulse for PCM/FM.

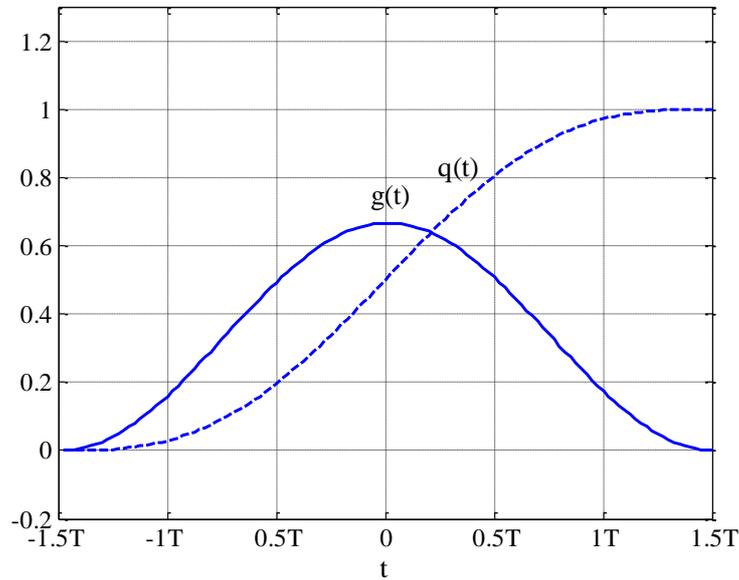

Figure 3: frequency pulse and corresponding phase pulse for ARTM CPM.

The baseband CPM signal $s(t;\alpha)$ can be expressed as follows [19]:

$$s(t;\alpha) = \exp\{j\varphi(t;\alpha)\}, \qquad (1)$$

where the instantaneous phase φ can be computed by

$$\varphi(t;\alpha) = \frac{\pi}{T}\int_{-\infty}^{t}\sum_{i=0}^{n}h_i\alpha_i g(\tau - iT)d\tau = \pi\sum_{i=0}^{n}h_i\alpha_i q(t - iT). \qquad (2)$$

In Equation 2, $\alpha_i$ represents $i^{th}$ information symbol and its value range is determined by the modulation order $M$; $h$ is the modulation index; $T$ is the duration of each symbol; the detailed modulation parameters for PCM/FM (Tier 0) and ARTM CPM (Tier 2) are given in Table 1; $g(t)$ is the frequency pulse function and $q(t)$ is the corresponding phase pulse function. The typical $g(t)$ and $q(t)$ are shown in Figure 2 and Figure 3 for PCM/FM and ARTM CPM respectively. In this paper, the additive white Gaussian noise (AWGN) channel is assumed, which is typical in the telemetry system. At the receiver, the received signal $r(t)$ can be written as follows:

$$r(t) = s(t; \alpha) + n(t), \tag{3}$$

where $n(t)$ is zero-mean complex AWGN vector, whose real part and imaginary part are uncorrelated with power spectral density $n_0$ in each signal dimension.

Table 1: The modulation parameters for PCM/FM and ARTM CPM.

|  | PCM/FM | ARTM CPM |
| --- | --- | --- |
| Modulation order ($M$) | 2 | 4 |
| Symbol values | $\alpha_i \epsilon \{-1, 1\}$ | $\alpha_i \epsilon \{-3, -1, 1, 3\}$ |
| Modulation index | $h=7/10$ | $h_i \in \{4/16, 5/16\}$ |
| Frequency pulse | Six-order Bessel ($L=3$) | 3 RC |

$L$ is the duration of pre-filter.

**Maximum likelihood sequence detection (MLSD)**

CPM is a modulation scheme with memory. That means the current modulation symbol is relevant to all the previous symbols. The Ref [23] indicates that the MLSD method is suitable to detect this kind of signal. As mentioned in section I, the MLSD criterion is to minimize the error probability of the whole sequence. When the received signal has the form of Equation 3, according to the MLSD criterion, the detector's task is to find an optimal estimation of source sequence, $\widetilde{\boldsymbol{\alpha}}$, to maximize the following likelihood function [24]:

$$\max_{\boldsymbol{\alpha}} L(\boldsymbol{\alpha}) = \exp\left\{-\frac{1}{n_0}\int |r(t) - s(t; \alpha(t))|^2 \, dt\right\}. \tag{4}$$

After logarithm operation on both sides of Equation 4 and simplification, the optimal estimation is equivalent to following solution [19]:

$$\widetilde{\boldsymbol{\alpha}} = arg \max_{\boldsymbol{\alpha}} \{Re[\int r(t)s^*(t; \alpha(t))dt]\}, \tag{5}$$

where $s^*(t; \alpha(t))$ represents the conjugate of $s(t; \alpha(t))$. The complexity of calculating Equation 5 will increase exponentially as the number of received symbols increases. According to Equation 2, at time $t=nT$, we have:

$$\int_0^{(n+1)T} r(t)s^*(t; \alpha)dt = \int_0^{nT} r(t)s^*(t; \alpha)dt + e^{-j\vartheta_{n-3}} \int_{nT}^{(n+1)T} r(t)e^{-j\theta(t;\alpha_n)} \, dt, \tag{6}$$

where $\vartheta_{n-3} = \pi \sum_{i=0}^{n-3} h_i \alpha_i$, and $\theta(t; \alpha_n) = \pi \sum_{i=n-2}^{n} h_i \alpha_i q(t - iT)$. The receiver structure characterized by Equation 6 is shown in Figure 4.

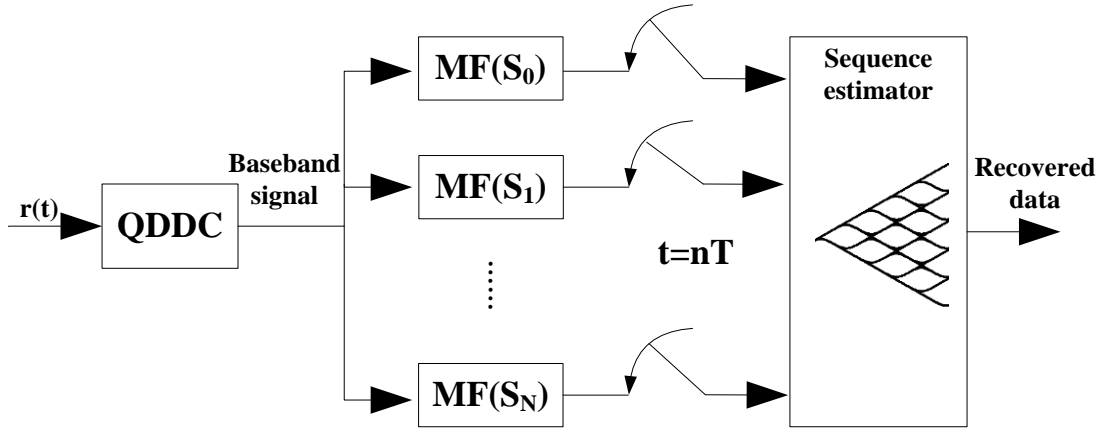

Figure 4: The MLSD detector for CPM using matched filter and VA.

In Figure 4, the QDDC module represents quadrature digital down conversion, MF is the matched filter, and the number of MF is $M^L$. The VA is used to complete the maximum likelihood sequence estimation. In VA, the state vector is defined as follows:

$$\boldsymbol{\sigma} = \{\vartheta_{n-3}, \alpha_{n-2}, \alpha_{n-1}\} \tag{7}$$

In Equation 6 the term $\int_0^{nT} r(t)s^*(t;\alpha)dt$ represents the partial path metric, and $e^{-j\vartheta_{n-3}} \int_{nT}^{(n+1)T} r(t)e^{-j\theta(t;\alpha_n)}dt$ represents the branch metric. It is clear that there are $pM^{L-1}$ states in the MLSD structure, where $p$ is the number of all possible phase state $\vartheta_{n-3}$, both $M$ and $L$ can be found in Table 1. According to [25], for the parameter $\vartheta_{n-3}$, there are 20 possible values $\{0, 7\pi/10, 2 \times 7\pi/10, \cdots, 19 \times 7\pi/10\}$ in PCM/FM and 32 possible values $\{0, \pi/16, 2\pi/16, \cdots, 31\pi/16\}$ in ARTM CPM respectively. The received signal is first performed the correlation operation with $M^L$ matched filters, then the matched filter outputs are rotated according to the values of the cumulative phases to obtain the branch metric. Finally the branch metric is added to the path metric of previous state, with this, the partial path metrics in the trellis is updated.

## Proposed low-complexity noncoherent maximum likelihood sequence detection

Although the MLSD algorithm can guarantee the minimum sequence error probability, the carrier's frequency and phase must be recovered first at the receiver. Otherwise it will lead to degradation in the performance. Unfortunately, in practical telemetry system the accurate phase of the carrier is difficult to acquire at the receiver, especially for the CPM signal. In this section how to detect the CPM signal with unknown carrier phase under the MLSD criterion is discussed. We deduce the detection criterion with unknown phase firstly, and then the low-complexity implementation structure is presented.

**Detection Criterion with Unknown Phase**

When the phase of carrier is unknown, the received signal can be expressed as follows:

$$r(t;\alpha) = s(t;\alpha)e^{jv} + n(t), \tag{8}$$

where $v$, is the initial phase, a uniformly distributed random variable taking values between 0 and $2\pi$, consequently its probability density function (PDF) is $f(v) = \frac{1}{2\pi}$.

Compare Equations 3 and 8 there is additional phase factor $e^{jv}$ in Equation 8. This makes the likelihood function for the received signal in Equation 8 different from that in Equation 3, and an expectation operation over the random initial phase must be implemented for the case of unknown carrier phase. As a result, the likelihood function is modified as follows:

$$L(\alpha) = \int_{-\pi}^{\pi} exp\left\{-\frac{1}{n_0}\int|r(t)e^{jv}-s(t;\alpha)|^2 dt\right\} f(v)dv = \int_{-\pi}^{\pi} L_1(\alpha)L_2(\alpha)f(v)dv, \qquad (9)$$

where $L_1(\alpha)$ and $L_2(\alpha)$ are defined respectively as follows:

$$L_1(\alpha) = exp\left\{-\frac{1}{n_0}\{\int|r(t)e^{jv}|^2 dt + \int|s(t;\alpha)|^2 dt\}\right\}, \qquad (10)$$

$$L_2(\alpha) = exp\left\{\frac{2}{n_0}\{Re[\int r(t)s^*(t;\alpha)dt]cosv + Re[\int j\cdot r(t)s^*(t;\alpha)dt]sinv\}\right\}. \qquad (11)$$

Since the CPM signal has a constant envelope, $L_1(\alpha)$ is a constant, and it is not difficult to infer that $L(\alpha)$ is the zero-order modified Bessel function [24]. Then we have:

$$L(\alpha) = CI_0(|\int r(t)s^*(t;\alpha)dt|), \qquad (12)$$

where $C$ is a constant. Because $I_0(x)$ is a non-decreasing function, the detection for PCM signals based on the ML criterion is equivalent to find the optimal estimation $\widetilde{\alpha}$ to satisfy following equation:

$$\widetilde{\alpha} = arg\max_{\alpha}\{|\int r(t)s^*(t;\alpha)dt|\} \qquad (13)$$

Compare Equations 5 and 13, it is clear that when the carrier phase is changed to be unknown the ML estimation converts to compute the module value of the integral term $\int r(t)s^*(t;\alpha)dt$, instead of its real part. From Equation 13, the module value of the integral term is irrelevant to the initial phase $v$, therefore the modulo operation can eliminate the effect of the initial phase uncertainty on the sequence estimation result, i.e., the solution of this formula can make the error probability of the sequence minimum when the initial phase is unknown. Besides, Equation 13 can also be similarly written as format of Equation 6, and we can adopt VA to complete sequence estimation. However, since the original VA is extremely complex, modulo operation will make solution more complex. As a result, how to solve the Equation 13 with the lowest possible complexity and keeping the desirable performance is the key. A low-complexity detection scheme is explored in the following part.

**Low-complexity Implementation Scheme**

As mentioned above, if we continue to adopt original VA in part II to solve the Equation 13, the complexity will be unacceptable for practical implementation. From the Equation s 6 and 13, it should be noticed that initial phase $v$ has no effect on the value of $|\int r(t)s^*(t;\alpha)dt|$. That can be explained as follows. At time $t=kT$, the state vectors of two paths are $\boldsymbol{\sigma}^{(1)} = \{\vartheta_{k-3}^{(1)}, \alpha_{k-2}^{(1)}, \alpha_{k-1}^{(1)}\}$ and $\boldsymbol{\sigma}^{(2)} = \{\vartheta_{k-3}^{(2)}, \alpha_{k-2}^{(2)}, \alpha_{k-1}^{(2)}\}$ respectively. When the equation $\vartheta_k^{(1)} - \vartheta_k^{(2)} = c$ is satisfied, where $c$ is a constant, the two paths will produce the same detection result. As a result, we do not need to calculate all possible values of $\vartheta_{n-3}$ in the detection structure given in section 2. And also there is the relationship:

$$\vartheta_{n-3} = \vartheta_{n-4} + \alpha_{n-3} h_i \pi. \tag{14}$$

It means $\vartheta_{n-3}$ is not an independent component of the state vector, and it is jointly determined by $\vartheta_{n-4}$ and $\alpha_{n-3}$. The phase item $\vartheta_{n-3}$ can be seen as a parameter, which is passed from the previous state to the current state, instead of an element of the state vector. In this sense, the state vector can be defined as follows:

$$\boldsymbol{\sigma}' = (\alpha_{n-2}, \alpha_{n-1}, \alpha_n | \vartheta), \tag{15}$$

Where $\vartheta$ is the phase corresponding to the current state, and it is named as survived phase determined by the phase of previous state and the survived path. The state transition is only determined by the information bits. As a result, when receiving a new symbol, only $M^L$ states need to be updated instead of $pM^{L-1}$, and also only $M^{L+1}$ branches need to be calculated instead of $pM^L$. Usually, existing the relationship $p \gg M$, consequently it will bring significant decrease in calculation complexity.

**Viterbi Algorithm with Phase Parameter**

When using state vectors defined in Equation 15, the VA for MLSD needs to be modified. Each state keeps not only the survived paths but also the survived phase, which is different from the original VA. Once one correlation value from the matched filter is received, the sequence estimator firstly rotates the phase of the branch metric (BM) according to the corresponding survived phase in previous state, then the rotated branch metric is added to corresponding survived path metric to create the new path metric (PM). After all states are updated, the sequence estimator will compare the branch metrics for each state to decide the survived path, and the phase value corresponding to the survived path is chosen as the survived phase of current state. Figure 5 shows the state transition diagram of the proposed method for PCM/FM and ARTM CPM.

From Figure 5, it is clear that the implementation structure is similar to MLSD in Figure 4 but with an essential difference when calculating the branch metric and updating the state information. Figure 6 depicts an example on how to calculate the branch metric and update the state information in detail. Assume the current state is $S_l$, and the previous state may be $S_m$ or $S_n$, and the corresponding survived phase and path metric are ($P_m$, $D_m$) or ($P_n$, $D_n$) respectively. After calculating the path metrics, the path with larger PM is chosen as the survived path, and the rotated phase corresponding to the survived path $P_m + h_i \alpha_m \pi$ or $P_n + h_i \alpha_n \pi$ is the survived phase, $P_l$, of the current state.

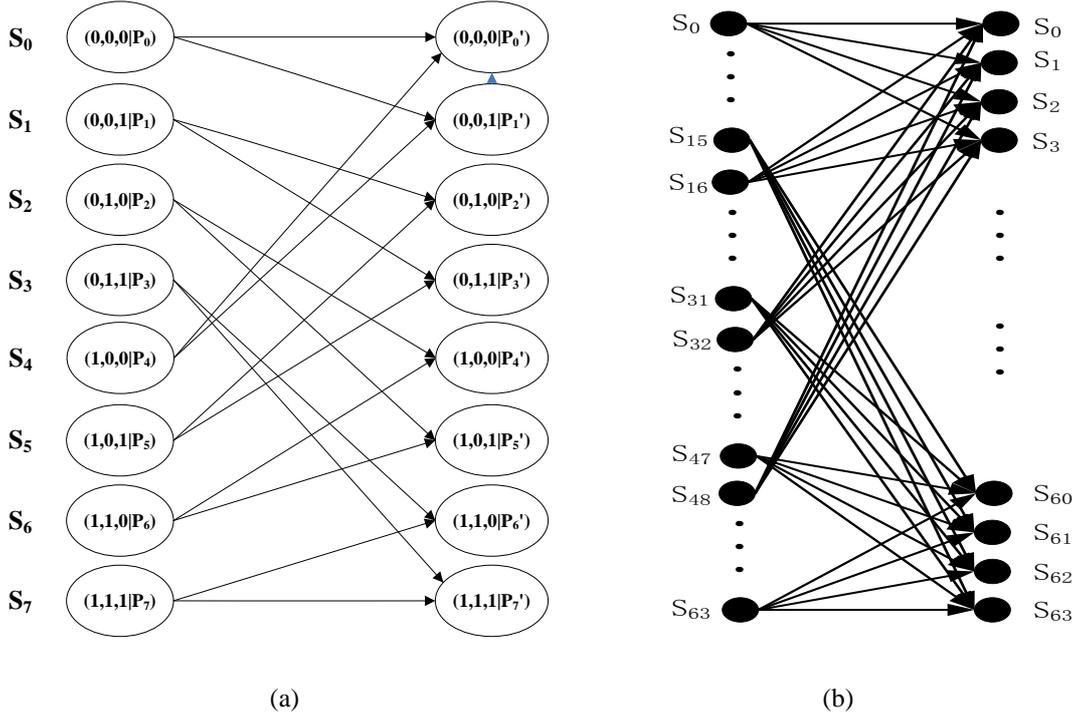

Figure 5: The state transition diagram. (a) PCM/FM. (b) ARTM CPM.

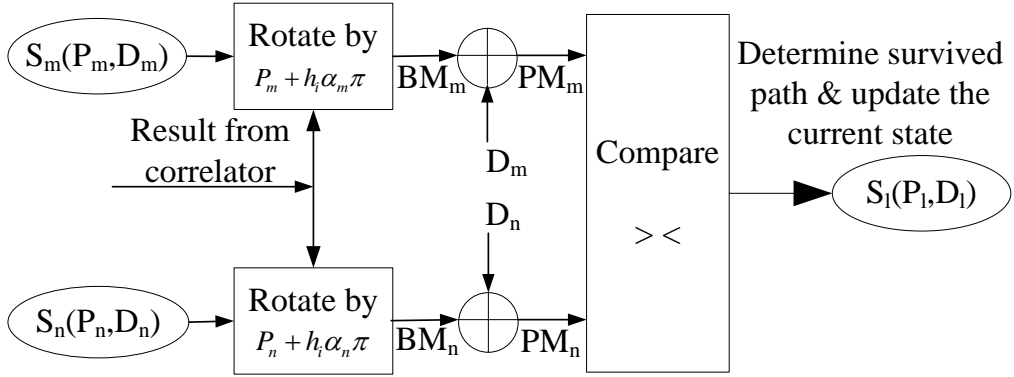

Figure 6: Calculation process of state transition.

The entire process of proposed scheme with VA is stated in Algorithm 1.

**Algorithm 1** Viterbi Algorithm for The Proposed Scheme
1: **for** $i$=1 to $N$ **do**
2:     calculate the output of MFs for the $i^{th}$ received symbol
3:     **for** $j$=0 to $M^{L}$-1 **do**
4:         calculate the previous states, $S_1(D_1,P_1), S_2(D_2,P_2),... S_M(D_M,P_M)$ according to state $j$
5:         calculate $PM_1, PM_2, ... PM_M$ according to Figure 6
6:         choose the path which has largest $PM$ as the survived path
7:         update state $j$
8:     **end for**
9: **end for**
10: trace back the survived path and output the decision sequence $\{\alpha_1, \alpha_2, \cdots \alpha_N\}$

As mentioned above, when the VA with phase parameter is used, since the number of states reduces dramatically, the complexity decrease compared to the original MLSD algorithm.

**Performance Analysis**

The performance of MLSD for CPM, characterized by bit error rate (BER) $P_b$, is mainly determined by the minimum distance, $d_l$, between sequences and the number of sequence pairs which have the minimum distance [19]:

$$P_b \leq \sum_{l=0}^{\infty} C_l\, Q\left(\sqrt{d_l^2 \frac{E_b}{n_0}}\right), \tag{16}$$

where $Q(x) = \frac{1}{\sqrt{2\pi}} \int_x^{\infty} e^{-u^2/2} du$, the parameter $C_l$ increases with the increase of the number of sequence pairs which have the minimum distance, and $E_b$ is the equivalent bit energy. For ARTM CPM using the MLSD method, it is sufficient to use the first two terms in Equation 16 to produce the approximation of $P_b$. When using the proposed VA, we assume two of the previous states, denoted by $\sigma'^{(1)}$ and $\sigma'^{(2)}$ (see Equation 15), which enter the current state. If $\alpha_{n-3}^{(1)} \alpha_{n-3}^{(2)} > 0$ is satisfied, some paths in the new phase trajectory trellis are probable to merge earlier, which makes these merged paths have near-minimum-distance compared to the original VA and corresponding terms cannot be neglected when calculating the $P_b$ in Equation 16. Consequently, the detection performance will decline. Apparently, for PCM/FM there are two branches for each state. The decision bits, $\alpha_{n-3}$, in Equation 14, which are used to calculate rotated phases of two different branches, are always of different sign, i.e., $\alpha_{n-3}^{(1)} \alpha_{n-3}^{(2)} < 0$. In this sense, the difference between two branches is large enough so that the paths are almost impossible to merge earlier. Therefore the detection performance has almost no loss in PCM/FM. However, for ARTM CPM there will be four branches for each state, and $\alpha_{n-3}$ in Equation 14 from different branches are possible to have the same sign, i.e., $\alpha_{n-3}^{(1)} \alpha_{n-3}^{(2)} > 0$. As a result, the performance will undergo a decrease in ARTM CPM when the proposed algorithm is used.

In order to overcome this problem, we can appropriately increase the number of survived paths in each state, i.e., each state will keep two survived paths or more. It should be noticed that even though more survived paths are kept, the complexity in proposed algorithm is still rather low. This is because that the complexity is linear with the number of survived paths and the number of states, and the detection performance can be guaranteed in high-order modulation so long as the number of survived paths increases slightly. That is to say the number of states decreasing is much larger than the number of path increasing in the proposed algorithm. Consequently, the modified VA can achieve an ideal trade-off between complexity and performance to satisfy different application requirement. However, in the following sections, if not specified, the proposed method keeps only one survived path.

## Experimental results

Simulations are carried out under standard complex AWGN channels at signal to noise ratios (SNRs) from 6 dB to 10 dB for 5-symbol BQCR-MSD [13], MLSD [19] and the proposed method in this paper. The PCM/FM signal and the ARTM CPM signal are modeled in the same way as that in the part of system model. The uncertain initial phase in received signal is modeled as a random variable with uniform distribution in the interval $[0, 2\pi)$ and it is regenerated for each data frame. The number and length of the data frame are $10^4$ and $10^3$, respectively for all methods. Table 2 and Table 3 provide the comparisons of computational complexity and storage requirement respectively for these three methods when receiving one symbol. The BER performances of PCM/FM and ARTM CPM are given in Figure 7 and Figure 8 respectively.

**Complexity Analysis**

For PCM/FM and ARTM CPM signals, the complexity is determined by the number of matched filters, $N_{mf}$, the number of survived path for each state, $N_p$, and the number of survived states, $N_s$. It should be noticed that $N_p$ takes value of 2 for ARTM CPM using the proposed method with two survived paths, otherwise $N_p=1$. We assume the oversampling rate of each symbol is $k$, that means there are $k$ sample values in one baseband symbol. When receiving one signal, the number of multiplications and the number of additions required, represented by $N_{mul}$ and $N_{add}$ respectively, are given as follows:

$$N_{mul} = 4N_{mf}k + 4N_s N_p M + 2N_s N_p M\delta, \quad (17)$$

$$N_{add} = N_{mf}(5k - 2) + 3N_s N_p M + N_s N_p M(1 + 2\delta), \quad (18)$$

where $\delta = 0$ for MLSD, and $\delta = 1$ for MSD and the proposed method. Typically, the oversampling rate $k$ takes the value of 4. It should be noticed that there is no state for the BQCR-MSD method, and $N_s N_p M = 128$.

Table 2: Calculation complexity comparison.

|  | BQCR-MSD | MLSD | | Proposed method | | |
|---|---|---|---|---|---|---|
|  | PCM/FM | PCM/FM | ARTM CPM$_1$ | PCM/FM | ARTM CPM$_1$ | ARTM CPM$_2$ |
| $N_{mf}$ | 8 | 8 | 64 | 8 | 64 | 64 |
| $N_s$ | —— | 80 | 512 | 8 | 64 | 64 |
| $N_{mul}$ | 896 | 768 | 9216 | 224 | 2560 | 4096 |
| $N_{add}$ | 912 | 784 | 9344 | 240 | 2688 | 4224 |

Subscript in ARTM CPM is the number of survived paths for each state.

Table 3: Storage requirement comparison.

|  | BQCR-MSD | MLSD | | Proposed method | | |
|---|---|---|---|---|---|---|
|  | PCM/FM | PCM/FM | ARTM CPM$_1$ | PCM/FM | ARTM CPM$_1$ | ARTM CPM$_2$ |
| Local signal | 64 | 64 | 512 | 64 | 512 | 512 |
| Rotation angle | 10 | 40 | 64 | 40 | 64 | 64 |
| Survived path | —— | 80$N$ | 512$N$ | 16$N$ | 128$N$ | 256$N$ |
| Survived phase | —— | 0 | 0 | 8 | 64 | 64 |

Storage unit is the capacity of one sample value, and $N$ is the trace back length of Virterbi algorithm.

As can be seen from Table 2, for PCM/FM, the number of multiplications in the proposed method is only about 29% of that in MLSD, and much less than that in BQCR-MSD. This is because that the multiplication operation only exists in matched filtering process and branch metric calculation. Compared to the MLSD, the number of states in the proposed method reduces to one tenth, which makes a significant decline in the number of branches metrics, thus leading to a considerable decrease in the multiplications. Besides, there is no branch calculation but relatively many matched filtering operations in BQCR-MSD, which leads to the most multiplications in BQCR-MSD. In addition, for PCM/FM, the number of addition in the proposed method only achieves 31% of MLSD and 26% of BQCR-MSD respectively.

For ARTM CPM, compared to the MLSD method, the number of states in the proposed method is approximate one eighth of the former, and both multiplication operation and addition operation in the proposed method with one survived path reduce by 70%. When the number of survived paths for ARTM CPM is doubled in the proposed method, the complexity increases by less than 60%, thus the proposed method still has much lower complexity compared with the other two methods. This verifies the previous analysis in Section III.

When it comes to Table 3, for PCM/FM, these three methods have the same number of local signals storage, since they use the same matched filters to obtain the correlation values. It is clear that BQCR-MSD needs less storage for rotation angle because its observation interval is shorter than the other two methods, and only a few phase values need to be stored. On the other hand, the storage for survived paths in the proposed method is only one fifth of that in MLSD. For ARTM CPM, both the traditional MLSD and the proposed method require same amount of storage in terms of local signal and rotation angle. However, due to the decline in number of states, the proposed method has a dramatic decrease in terms of storage for survived path compared to traditional MLSD, decreasing by 75%. When keeping two survived paths in ARTM CPM, the storage in the proposed method is still much lower compared to traditional MLSD due to significant decline in the number of states. What's more, the storage for additional survived phase in the proposed method is so low that it can be neglected.

From above analysis and comparison, the proposed method shows obvious superiority in terms of computation complexity and storage requirement compared to the traditional MLSD method. In addition, the proposed method greatly exceeds BQCR-MSD in terms of calculation complexity.

**BER Performance**

Even though the computational complexity decreases significantly, the performance of the proposed algorithm is almost the same as the conventional MLSD.

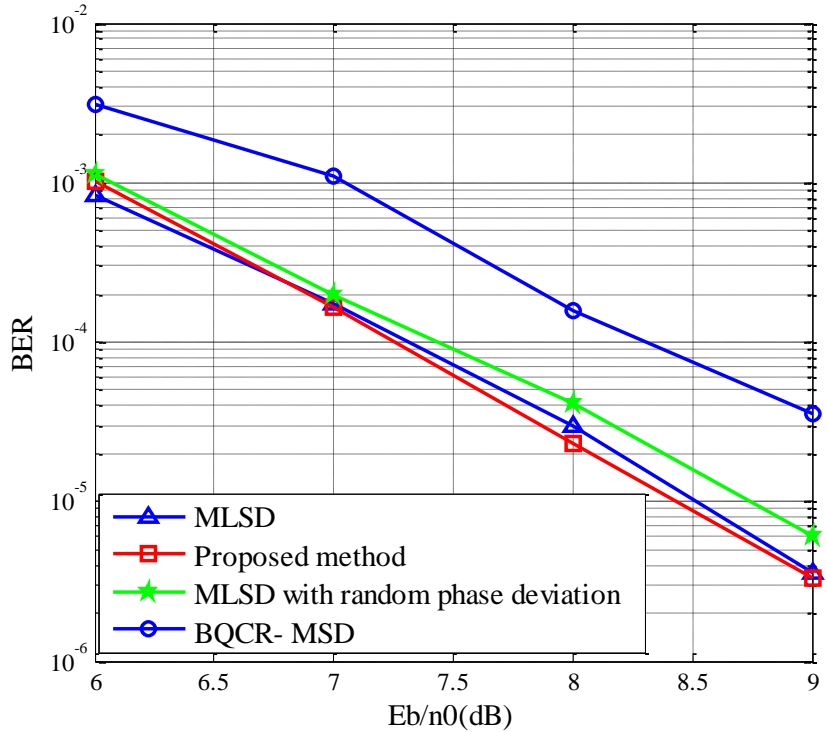

Figure 7: BER performance comparison for PCM/FM.

As is shown in Figure 7, for PCM/FM, in terms of BER performance, the proposed method is 1dB better than the BQCR-MSD and almost the same as the conventional MLSD with accurate phase recovery. When there is phase deviation varying from 0 to $2\pi$ randomly between the local signal and the received signal, the proposed method achieves lower BER compared to the traditional MLSD method, especially in high SNRs region.

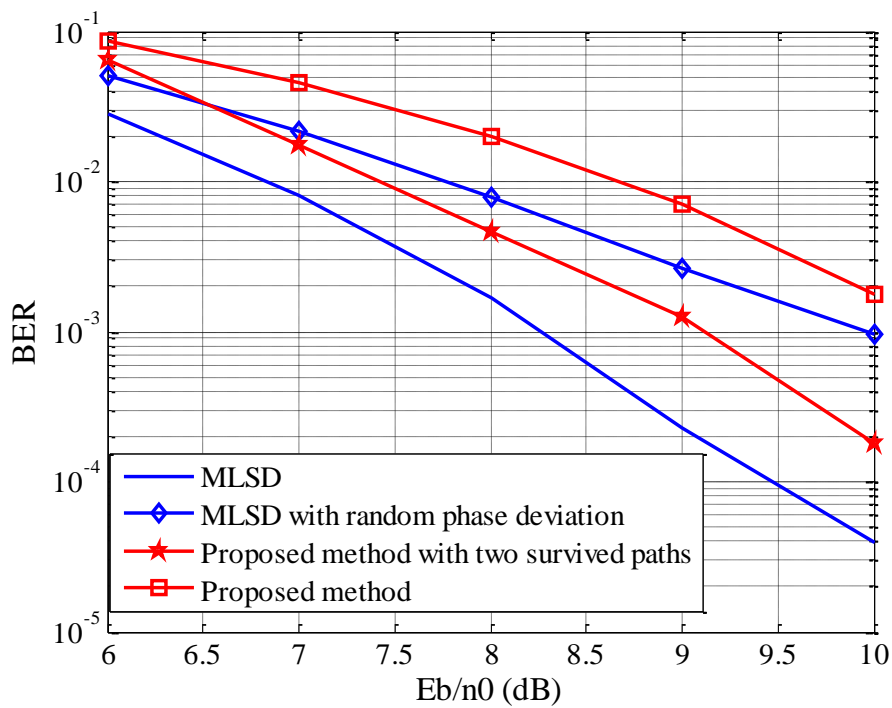

Figure 8: BER performance comparison for ARTM CPM.

From Figure 8, the traditional MLSD method with accurate carrier phase provides the best BER performance. However, when the phase deviation exists, the performance of the traditional MLSD undergoes a significant decline. The BER in the proposed method with one survived path is high, but the proposed method with two survived paths improves the BER performance significantly, and exceeds the traditional MLSD method with phase deviation. What's more, the gain grows with the increase of SNR. Although the proposed method with two survived paths is inferior to traditional MLSD with accurate carrier phase, considering the difficulty in accurate carrier recovery and high complexity in the traditional MLSD method, the proposed method is more practical.

From the experiment results and discussion, it is clear that the proposed method reduces the complexity significantly and keeps reasonable detection performance.

## Conclusions

In this paper, a low-complexity noncoherent MLSD method for PCM/FM and ARTM CPM is proposed and a corresponding implementation structure is also developed. It is proved that the proposed method does not need accurate carrier phase recovery which is more suitable for practical telemetry systems than the traditional MLSD method. The computational complexity in proposed scheme can be significantly reduced by optimizing the trellis states. Both the simulation results and theoretical analysis show that the BER performance in the proposed method is almost the same as the traditional MLSD with perfect carrier phase recovery, and greatly outperforms the BQCR-MSD. Moreover, the proposed method achieves better detection performance than traditional MLSD with practical carrier phase deviation.

Apparently, through redefining the state vector in VA trellis, the proposed method reduces the number of states and thus decreases the complexity. This does not conflict with the approaches of reducing complexity discussed in Section I. In other words, we can further reduce the complexity by applying other approaches on the basis of the proposed method in this paper.

## Conflicts of Interest

The authors declare that there is no conflict of interest regarding the publication of this paper.

## Acknowledgments

This research was supported by the National Natural Science Foundation of China (no. 31700479 and no. 91438116), and Fundamental Research Funds for the Central Universities (no. BLX201623).